\documentclass[prb,twocolumn,showpacs,preprintnumbers,amsmath]{revtex4}
\usepackage{dcolumn}
\usepackage{bm}
\usepackage{graphicx}
\usepackage{bm}
\begin{document}
\title{Real Space Visualization of Thermomagnetic Irreversibility within Supercooling and Superheating Spinodals in $Mn_{1.85}Co_{0.15}Sb$ using Scanning Hall Probe Microscopy}
\author{Pallavi Kushwaha, Archana Lakhani, R Rawat, A Banerjee and P Chaddah}
\affiliation{UGC-DAE Consortium for Scientific Research\\University Campus, Khandwa Road\\
Indore-452001, India.}
\date{\today}

\begin{abstract}
Phase coexistence across disorder-broadened and magnetic-field-induced first order antiferromagnetic to ferrimagnetic transition in polycrystalline $Mn_{1.85}Co_{0.15}Sb$ has been studied mesoscopically by Scanning Hall Probe Microscope at 120K and up to 5 Tesla magnetic fields. We have observed hysteresis with varying magnetic field and the evolution of coexisting antiferromagnetic and ferrimagnetic state on mesoscopic length scale. These studies show that the magnetic state of the system at low field depends on the path followed to reach 120 K. The low field magnetic states are mesoscopically different for virgin and second field increasing cycle when 120 K is  reached by warming from 5K, but are the same within measurement accuracy when the measuring temperature of 120K is reached from 300K by cooling.
\end{abstract}


\maketitle
\section {Introduction}
First order magnetic transitions have been of extensive scientific interest in recent years. The interest in these systems arise due to their technological importance like giant magnetoresistance, magnetocaloric effect, magnetic shape memory effect etc. as well as their fundamental importance to understand various interesting phenomena like phase separation, metastability, glass like magnetic state etc. Quench disorder in a system can lead to spread of local transition temperature resulting in the broadening of a first order transition \cite{Imry}.  This broadening gives rise to coexistence of competing phases in the transition region. The metastability of coexisting phases within (and below) the supercooling and superheating spinodals has been of wide interest and actively pursued in a wide variety of systems like systems showing metal insulator transitions \cite{Toku, Shar, Wang}, multiferroics \cite{Xu}, intermetallics \cite{Roy} etc. The understanding of magnetic first order transition (due to easy control of magnetic field (H) and temperature (T) ) also has implication to wider class of systems where first order transition plays a role (like glass transition where pressure and quenching rate are sometimes difficult to control). It has been argued that the glass like metastable states resulting from the slow dynamics of the transition are different from the metastable states which arise due to supercooling and superheating near the first order transition \cite{Kuma}. Both kind of metastable states can show seemingly similar features in some of their physical properties e.g. open hysteresis loop in isothermal R-H measurement. However, it has been shown recently \cite{Palv} that such open hysteresis loop due to supercooling and superheating will be observed only for T within these spinodals and will be observed only during cooling or only during heating, depending on the sign of the slope of transition band in H-T space. Mesoscopic investigation by Scanning Hall Probe Microscopy  (SHPM) has shown coexisting antiferromagnetic (AFM) and ferromagnetic (FM) phase around critical field in doped $CeFe_2$ and $Gd_5Ge_4$ \cite{Roy1, Moore}. For the T chosen in these studies, it did not matter whether the measurement temperature is reached by cooling or by warming. Here we present real space magnetic imaging study by SHPM along with magnetization and resistivity measurement of $Mn_{1.85}Co_{0.15}Sb$ to show that field induced transition for T (=120 K) lying between supercooling and superheating spinodal depends on the path followed to reach the measurement temperature.

Doped $Mn_2Sb$ shows first order antiferro (AFM) to ferrimagnetic (FRI) transition at low temperature \cite{Beck}. Below transition temperature ($T_N$) AFM to FRI transition can be induced with the application of magnetic field. When $T_N$ is shifted to lower temperature, these systems show anomalous magnetic behavior \cite{Bart, Zhan}. We have addressed some of these anomalous behavior in our magnetotransport studies of Co doped $Mn_2Sb$ \cite{Palv}. In these studies we have shown that anomalous thermomagnetic irreversibilities at low temperature are a result of critically slow dynamics of the transition and these are different from the seemingly similar irreversibility that arise due to supercooling and superheating.

\maketitle\section{Experimental Details}

$Mn_{1.85}Co_{0.15}Sb$ sample used in the present study is taken from same ingot which has been used for earlier resistivity/magnetoresistance studies \cite{Palv}. Resistivity/magnetoresistance measurement were performed using home made resistivity setup inside Oxford magnet system. Magnetization measurement were performed using VSM option of PPMS. 
Magnetic imaging was carried out using Scanning Hall Probe Microscope from NanoMagnetics Instruments, U.K. The microscope incorporates a chip sensor, which consists of a 1-micron size square Hall sensor integrated adjacent to a tunneling tip. The tunneling tip is used for bringing the Hall sensor in close proximity to sample surface. The sensor chip is aligned with a small angle ($\approx 1^0 $) to keep tunneling tip closer to sample surface than Hall sensor. Magnetic imaging is carried out by scanning the Hall sensor over the sample surface while simultaneously measuring the Hall voltage, which is proportional to perpendicular component of the magnetic field at the surface. In the present study we have carried out magnetic imaging in lift off mode. In this mode sample surface is reached by finding tunneling current. After finding the sample surface tip is retracted few hundred nanometers, called lift off, and scanning is performed at this constant height. For low temperature and high field measurements this insert is placed inside 9-Tesla superconducting magnet (American Magnetics) system supported on a two-stage vibration isolation stage. 
Approximately 6mm diameter and ~2 mm thick sample is polished to mirror finished surface for SHPM imaging. All the images in the present study are of $27 \mu m \times 27 \mu m$ scan area and pixel size $128 \times 128$, which were scanned at $0.3 \mu m$ lift off and $5 \mu m/sec$ scan speed.  
All the measurements were carried out as a function of Magnetic field at 120 K for two protocols; (i) Sample is cooled from 300 K to 120K (i.e reached by cooling) in zero field and (ii) sample is cooled to 5 K and then heated back to 120 K (i.e. reached by warming)in zero field. 

\maketitle\section{Results and discussion}

Figure 1 [a]and [b] show the temperature dependence of resistivity $(\rho)$ in zero magnetic field and magnetization (M) in 0.1 Tesla magnetic field respectively for cooling and then warming cycle. AFM (higher resistivity and lower magnetization) to FRI (lower resistivity and higher magnetization) transition is
visible as a sharp decrease in resistivity (increase in magnetization) with increasing temperature and shows a hysteresis of ~10K between heating and cooling cycle. The slightly lower transition temperatures obtained from magnetization measurement are in accordance with magnetic field dependence of $T_N$. Beside the 10 K hysteresis, transition is broad for both cycles during cooling as well as warming. This is expected for substitutional alloys where inherent chemical disorder can result in distribution of local transition temperature on the length scale of correlation length \cite{Imry}. The spread in local transition temperature result in a band of transition in H-T space and two phases (here FRI and AFM) can coexist within this band. Therefore, this broadening of first order transition makes this compound suitable to study the coexistence of phases  and their evolution with magnetic field. The schematic of $(H^*, T^*)$ and $(H^{**}, T^{**})$ for AFM to FRI transition is shown in the inset of figure 1 [a]. For the sake of simplicity $(H^*, T^*)$ and $(H^{**}, T^{**})$ spinodals are shown well separated in contrast to overlapping bands actually observed in the present system. Here, isothermal measurement were carried out along path QS for two conditions, viz when the point Q is reached (i) by following path PQ (cooling from $T > T^{**}$) and (ii) by following path RQ (heating from $T < T^{*}$). 

\begin{figure}[t]
	\begin{center}
	\includegraphics[width=8.5 cm]{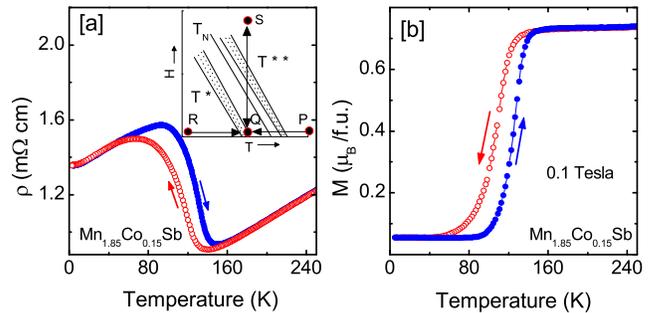}
	\end{center}
	\caption{\textbf{[a]} Resistivity in zero field and \textbf{[b]} Magnetization in 0.1 Tesla magnetic field as a function of temperature measured during cooling and subsequent warming for $Mn_{1.85}Co_{0.15}Sb$. Inset shows the schematic of supercooling $(H^*, T^*)$ and superheating $(H^{**}, T^{**})$ spinodal in H-T space. Isothermal measurements presented in Figure 2 (Figure 3) were carried out along path QS when point Q is reached by cooling (warming) following path PQ (RQ).}
	\label{Figure1}
\end{figure}
  
  Figure 2 shows some of the representative SHPM images as a function of magnetic field with increasing and then subsequent decreasing field at 120K when reached from 300 K (i.e. reached by cooling). All the images, shown in figure 2,  are plotted on same scale after subtracting the applied magnetic field. The labels on these images are marked on corresponding $\rho - H$ and $M-H$ curves (plotted in the middle row of figure 2) to correlate these results. Image (a) of figure 2, taken at 0.5 Tesla shows inhomogeneous magnetic state where both FRI (blue) and AFM (red) phases co-exists. The image contrast remains almost same with further increase in magnetic field to 1 Tesla, image (b), which is consistent with almost constant $\rho$ and $M$ between these field values.  At 2 Tesla image (c) shows increased FRI fraction and much smaller AFM fraction indicating a field induced AFM to FRI transition. Further increase in magnetic field to 4 Tesla results in homogeneous FRI state (image (d)). On reducing magnetic field from 5 Tesla to 2 Tesla image (e) shows almost homogeneous FRI phase in contrast to field increasing cycle where we observed coexisting FRI and AFM phase image (c). However image (f) taken at 1 Tesla during field decreasing cycle, shows inhomogeneous magnetic state which is similar to image (c) observed during field increasing cycle at 2 Tesla. This irreversibility is consistent with the first order nature of the field induced magnetic transition. Both resistivity value as well as magnetization value are identical for point `c' and point `f' as shown in bottom graphs. As the magnetic field reaches 0.5 Tesla the image (g) resembles image (a) taken during field increasing cycle for same field value i.e. the magnetic state of the system is same before and after the application of magnetic field at low field. This is consistent with the $\rho - H$ and $M-H$ curves, where zero field resistivity is found to be same before and after the application of magnetic field and virgin curve (curve taken during first field increasing cycle) overlaps with the envelope curve (taken during second field increasing cycle). 

\begin{figure}[t]
	\begin{center}
	\includegraphics[width=8.5 cm]{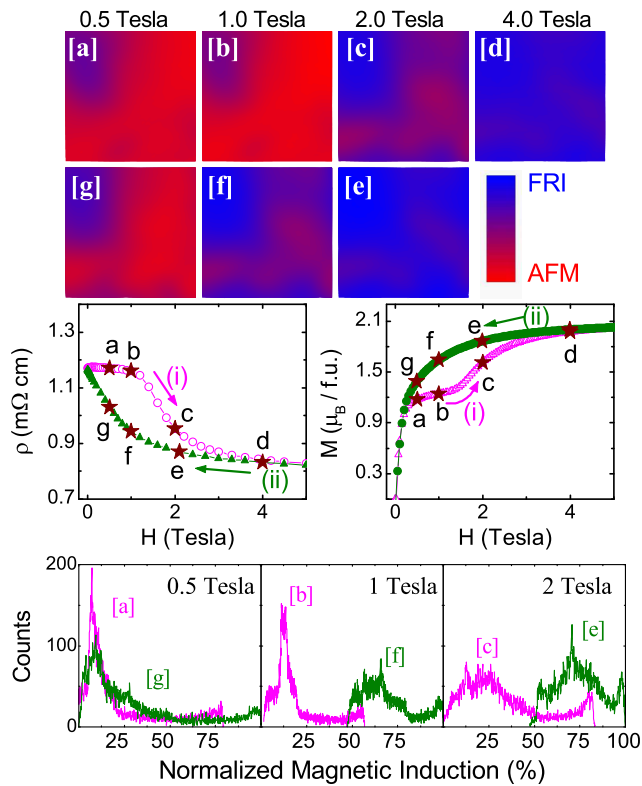}
	\end{center}
	\caption{\textbf{[a]-[g]} SHPM Images of $Mn_{1.85}Co_{0.15}Sb$ as a function of magnetic field at 120 K (reached by cooling from 300 K) along with corresponding resistivity and magnetization curve. Scan area is $27 \mu m \times 27 \mu m$ and image label corresponds to respective point in resistivity and magnetization curves. Bottom row shows the histograms of magnetic images at labeled field values for increasing and decreasing field cycle. Inhomogeneous magnetic state and similar magnetic state after field cycling at low field is highlighted along with characteristic hysteresis associated with magnetic field induced first order transition}
	\label{Figure2}
\end{figure} 

The weak contrast in the images arises due to bulk sample (thickness~2mm), whereas the observed phase separation is on the length scale of few $\mu m$. To demonstrate magnetic inhomogeneity more clearly, histograms of magnetic field distribution  are plotted in the bottom row corresponding to magnetic images shown in same figure. For the sake of comparison, field window for histogram calculation as well as vertical scale  are kept same in all the plots. For 0.5 Tesla both the curves (curve `a' and `g') are almost identical with slightly higher FRI phase for curve `g'. The magnetization at point `g' is only slightly higher than in point `a'. At 1 Tesla, curve `b' and curve `f', indicate entirely different magnetic field distribution on sample  surface during field increasing and decreasing cycle. Similar to 1 Tesla, we observe entirely different histograms corresponding to image (c) and image (e) taken at 2 tesla. 
 
\begin{figure}[b]
	\begin{center}
	\includegraphics [width=8.5 cm]{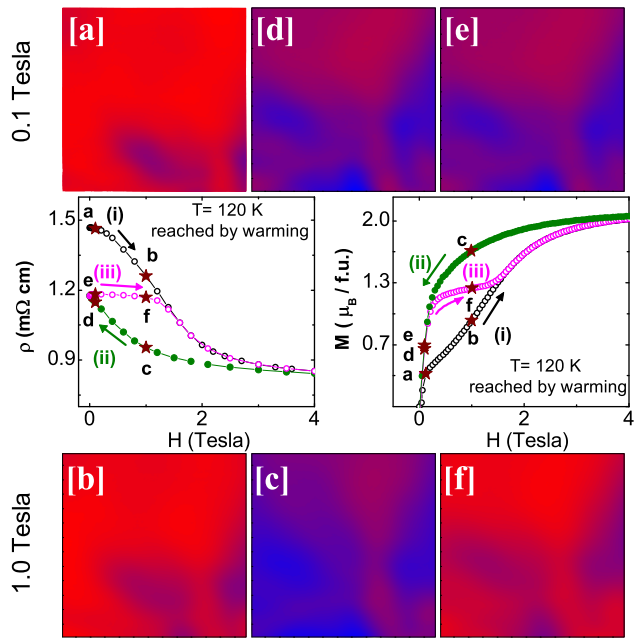}
	\end{center}
	\caption{\textbf{[a]-[f]} SHPM Images of $Mn_{1.85}Co_{0.15}Sb$ at 0.1 and 1.0 Tesla taken at 120 K (reached from 5 K) along with corresponding resistivity and magnetization curve. Scan area is $27 \mu m \times 27 \mu m$ and magnetic scale is same for each row separately.}
	\label{Figure3}
\end{figure} 

We repeated similar measurement under identical condition at 120K, when reached by warming from 5K under zero field condition. This experiment also shows a field induced AFM to FRI transition with varying field and associated irreversibility. However, our main interest is to study the state of the system at zero field before and after field cycling i.e. across virgin and envelope curve. Therefore in Figure 3, we show resistivity and magnetization data along with only two sets of images; one taken at 0.1 Tesla and other taken at 1 Tesla. For each set, images were taken during first field increasing cycle i.e. virgin curve , field decreasing cycle and second field increasing cycle. These curves are labeled as (i), (ii) and (iii) in $\rho - H$ and $M-H$ plots along with markers at which magnetic images were taken. To compare the SHPM images at constant field, magnetic scale is kept same for each set of images separately but varied for different magnetic field. As can be seen in top row (0.1 Tesla); image (a) is distinctly different from other two images (d) and (e) which are identical. Image (a) shows almost homogeneous AFM state  whereas other two images shows coexisting FRI and AFM states. This is consistent with the $\rho - H$ curve where point `d' and `e' have almost same resistivity but much smaller compared to point `a'. In case of $M-H$ also, M is same for point `e' and `d' and smaller for point `a'. This is in contrast to figure 2 (measured during cooling) where magnetic state of the system at low field is identical before and after the application of magnetic field. Images for 1 Tesla (bottom row) show that magnetic state of the system are more similar during curve (i) and curve (iii) (image (b) and image (f)) compared to that measured during curve (ii) (image (c)). Image (c), taken during field reducing cycle, has much larger FRI phase fraction compared to image (f) taken during field increasing cycle at same field value. This is consistent with $\rho - H$ curve where point `b' and point `f' have similar resistivity values compared to point `c'. A closer inspection of these images show higher FRI fraction in image (f) compared to image (b). 

\begin{figure}[h]
	\centering
	\includegraphics [width=8.5 cm]{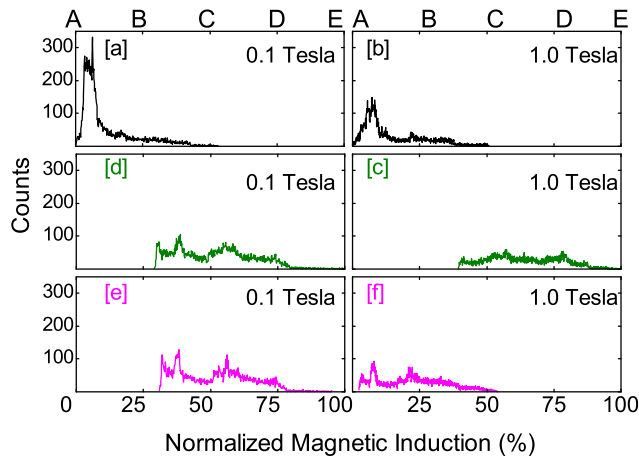}
	\caption{Histograms of SHPM Images of $Mn_{1.85}Co_{0.15}Sb$ at 0.1 Tesla and 1.0 Tesla for virgin curve, field decreasing cycle and second field increasing cycle at 120 K (reached by warming from 5 K). These curves highlight that the histogram corresponding to second field increasig cycle are distincly different from first cycle.}
	\label{Figure4}
\end{figure}
Histograms of magnetic images at low field are shown in figure 4 for 0.1 and 1.0 Tesla magnetic field for all the three cycles. Similar to figure 2, field window chosen for histogram calculation and vertical scale are kept same for all the figures. At 0.1 Tesla, histogram corresponding to image (a) has sharp peak and has distinctly different field distribution compared to curve (e).In case of histogram (a) about $87 \%$ of scanned region have magetic induction in the range AB where magnetic induction is less than $25 \%$ of the total scale, indicating almost homogenous AFM state for image (a). Whereas for histogram (d) and (e) more than $90 \%$ region has magnetic induction in the range BD ($25- 75 \%$ of total scale). Even at 1 Tesla there is a difference in magnetic field distribution for curve (b) and (f) though less drastic compared to that observed in 0.1 Tesla. Here also, more than $81 \%$ region for histogram (b) have magnetic induction in the range AB (0-25 \% of total scale) comapared to only $\approx60 \%$ regions for histogram (f) in the same range. It shows that there is more FRI phase fraction during second field increasing cycle at same field value compared to that observed during virgin curve. These observation show that magnetic state of the system changes from almost homogenous AFM state  to coexisting AFM and FRI state after field cycling at 120 K (when reached by warming). This explains the origin of open hysteresis loop in $\rho-H$ and virgin curve lying outside envelope curve in $M-H$ as well as $\rho-H$ measurements.

\maketitle\section{Conclusions}

The SHPM images of $Mn_{1.85}Co_{0.15}Sb$ at 120 K show almost homogenous AFM state at low field when 120 K is reached by warming in contrast to coexisting AFM and FRI state when reached by cooling i.e. the magnetic state of the system on mesoscopic length scale depends on the path followed to reach the measuerement temperature. Almost homogenous low field AFM state during warming is converted to coexisting AFM and FRI state on mesoscopic length scale after isothermal field cycling.  These studies provide the origin of open hysteresis loop observed in $\rho - H$ and virgin curve lying outside the envelope curve in $\rho - H$ and $M - H$ measurements observed during warming only. Similar studies on frozen glassy magnetic states will provide further insight on phase seperation and metastability.

\maketitle\section{Acknowledgments}

We acknowledge DST, India for funding of PPMS-VSM used in present study and Kranti Kumar for his help in magnetization measurement.

\end{document}